\documentclass[
twocolumn,
superscriptaddress,
amsmath,amssymb,
aps,
prl,
floatfix,
]{revtex4-2}

\usepackage{lipsum}
\usepackage{physics}
\usepackage{graphicx}% Include figure files
\usepackage{dcolumn}% Align table columns on decimal point
\usepackage{bm}% bold math
\usepackage[breaklinks=true,colorlinks,citecolor=blue,linkcolor=blue,urlcolor=blue]{hyperref}% add hypertext capabilities
\usepackage{titlesec}
\usepackage[caption=false]{subfig}

\renewcommand{\paragraph}[1]{\textit{#1}---}

\usepackage[capitalise]{cleveref}

\newcommand{\phantomsubcaptionlabel}[1]{\refstepcounter{subfigure}\label{#1}}

\newcommand{\rme}{{\rm e}}
\newcommand{\rmd}{{\rm d}}
\newcommand{\rmi}{{i}}

\begin{document}

\title{Floquet Dissipative Phase Transitions}% Force line breaks with \\

\author{Alberto Mercurio}
\email{alberto.mercurio@epfl.ch}
\affiliation{Institute of Physics, Ecole Polytechnique Fédérale de Lausanne (EPFL), CH-1015 Lausanne, Switzerland}
\affiliation{Center for Quantum Science and Engineering, Ecole Polytechnique Fédérale de Lausanne (EPFL), CH-1015 Lausanne, Switzerland}

\author{Vincenzo Macrì}
\affiliation{Dipartimento di Fisica ``A. Volta”, Università di Pavia, Via Bassi 6, 27100 Pavia, Italy}

\author{Filippo Ferrari}
\affiliation{Institute of Physics, Ecole Polytechnique Fédérale de Lausanne (EPFL), CH-1015 Lausanne, Switzerland}
\affiliation{Center for Quantum Science and Engineering, Ecole Polytechnique Fédérale de Lausanne (EPFL), CH-1015 Lausanne, Switzerland}

\author{Lorenzo Fioroni}
\email{lorenzo.fioroni@epfl.ch}
\affiliation{Institute of Physics, Ecole Polytechnique Fédérale de Lausanne (EPFL), CH-1015 Lausanne, Switzerland}
\affiliation{Center for Quantum Science and Engineering, Ecole Polytechnique Fédérale de Lausanne (EPFL), CH-1015 Lausanne, Switzerland}

\author{Vincenzo Savona}
\affiliation{Institute of Physics, Ecole Polytechnique Fédérale de Lausanne (EPFL), CH-1015 Lausanne, Switzerland}
\affiliation{Center for Quantum Science and Engineering, Ecole Polytechnique Fédérale de Lausanne (EPFL), CH-1015 Lausanne, Switzerland}

\date{\today}% It is always \today, today,

\begin{abstract}
Dissipative phase transitions (DPTs) are traditionally characterized through the spectrum of a time-independent Liouvillian superoperator.
However, this definition does not apply to time-periodic (Floquet) systems that cannot be exactly recast as time-independent problems.
Here, we develop a general framework to characterize DPTs in time-periodic open quantum systems through the spectrum of the Floquet propagator.
We first study driven-dissipative Kerr resonators, known to display a DPT, showing that counter-rotating terms in the drive shift the critical point and significantly change the time scales associated with the transition.
We then investigate DPTs in the driven quantum Rabi model and its time-independent approximation, the driven Jaynes-Cummings model, finding that the Rabi model exhibits distinct critical features as the ultrastrong coupling regime is approached.
Moreover, our Floquet analysis unveils the disappearance of the DPT in the deep strong coupling regime, due to light-matter decoupling.
Our approach sets the stage for the study of dissipative criticality in a broad class of time-dependent open quantum systems.
\end{abstract}

\maketitle

\paragraph{Introduction}
The study of periodically-driven (Floquet) quantum systems is at the forefront of non-equilibrium physics, unveiling a plethora of emergent phenomena absent in time-independent systems.
Among these, Floquet topological phases~\cite{cayssol_floquet_2013, rudner_anomalous_2013, rechtsman_photonic_2013, wang_observation_2013, jotzu_experimental_2014,  maczewsky_observation_2017, rudner_band_2020, wintersperger_realization_2020} and Floquet engineering of quantum materials~\cite{kitagawa_transport_2011, goldman_periodically_2014, bukov_universal_2015, holthaus_floquet_2016, oka_floquet_2019, mciver_light-induced_2020, weitenberg_tailoring_2021, park_steady_2022, merboldt_observation_2025, Akbari2025FloquetEngineeringQuantum, baykusheva_quantum_2026} have attracted significant attention.

With the advent of quantum technologies, understanding the interplay between coherent driving and the dissipation processes ubiquitous in quantum hardware has become increasingly important.
In several instances, the explicit time dependence can be effectively removed by retaining only the nearly resonant terms and neglecting the more rapidly oscillating ones, the counter-rotating terms, an approach known as the rotating wave approximation (RWA).
In many other cases, however, the intrinsic time dependence can neither be neglected nor simplified.
This is, for example, the case for protocols relevant to quantum computing, such as qubit control~\cite{gandon_engineering_2022, nguyen_programmable_2024, sun_logical_2025} and readout~\cite{shillito_dynamics_2022, cohen_reminiscence_2023, dumas_measurement-induced_2024, kurilovich2025highfrequencyreadoutfreetransmon, mencia2025raisingcavityfrequencycqed}, and to quantum simulation, including the engineering of synthetic lattices \cite{ozawa_topological_2019, meier_exploring_2019, arnal_chaos-assisted_2020, sridhar_quantized_2024} and parametric processes~\cite{lellouch_parametric_2017, dutt_single_2020, shan_giant_2021, goldman_floquet-engineered_2023, kiselev_inducing_2024, peyruchat_landauzener_2025}.

Dissipative phase transitions (DPTs), i.e. non-analytic changes in the steady state of an open quantum system \cite{minganti_spectral_2018}, are among the most striking phenomena in open many-body dynamics.
Beyond their fundamental importance, DPTs are attracting interest for their potential to enhance the performance of superconducting qubits~\cite{gravina_critical_2023, labaymora2025chiralcatcodeenhanced} and quantum sensing schemes~\cite{garbe_critical_2020, di_candia_critical_2023, alushi_optimality_2024, beaulieu_criticality-enhanced_2025, alushi_collective_2025, mihailescu_critical_2026}.
Although DPTs have been extensively investigated in time-independent open systems, their manifestation in time-periodic open setups \cite{sierant_dissipative_2022} remains largely unexplored.

In this paper, we address this question by introducing a general framework to characterize DPTs in Floquet open quantum systems through the spectrum of the Floquet propagator.
We apply it to paradigmatic systems in quantum optics: driven-dissipative nonlinear oscillators and the quantum Rabi model, which describes light-matter interaction at arbitrary coupling strengths.
Our results show that intrinsic time dependence can modify dissipative criticality both quantitatively and qualitatively, setting the stage for the study of phase transitions in more complex many-body time-periodic open quantum systems.

\begin{figure*}[t]
    \centering
    \includegraphics{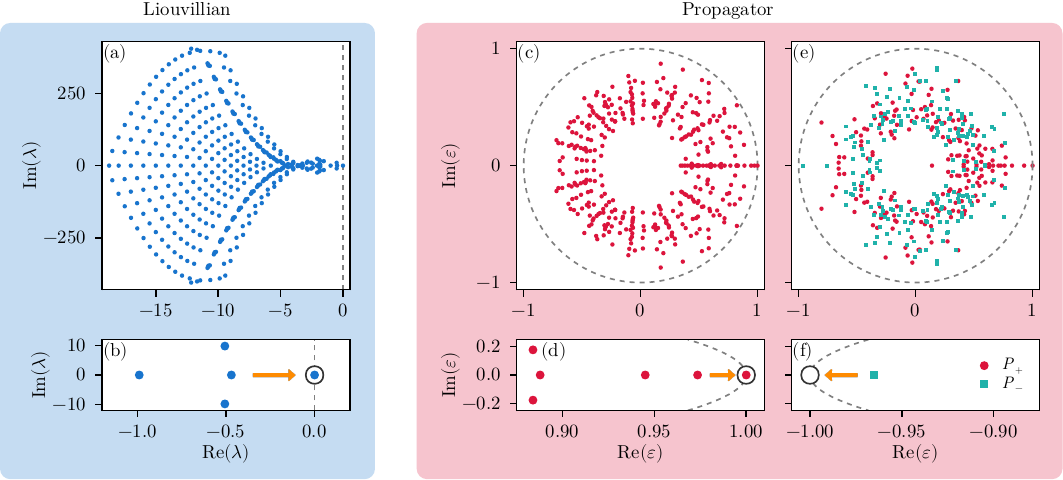}
    \caption{Spectral signatures of DPTs in time-independent and Floquet open systems. (a,b) Liouvillian picture, where criticality is marked by the closure of the Liouvillian gap, $\mathrm{Re}(\lambda_1)\to 0$. (c-f) Floquet picture, where the one-period propagator eigenvalues $\varepsilon_j$ lie within the unit circle. (c,d) Criticality is determined through the closure of the gap, $|\varepsilon_1|\to 1$. (e,f) With weak symmetry breaking (here $\mathbb{Z}_2$), the propagator splits into sectors ($P_+$, $P_-$), and the sector-resolved leading eigenvalues approach the corresponding symmetry roots of unity, here $\Re(\varepsilon_{1, 0}) \to +1$ and $\Re(\varepsilon_{0, 1})\to -1$. Notice that the angular position of the eigenvalues depends on the frame, as discussed in the End Matter.}
    \label{fig-eigenvalues_sketch}
    {\phantomsubcaptionlabel{fig-eigenvalues_sketch-a}}
    {\phantomsubcaptionlabel{fig-eigenvalues_sketch-b}}
    {\phantomsubcaptionlabel{fig-eigenvalues_sketch-c}}
    {\phantomsubcaptionlabel{fig-eigenvalues_sketch-d}}
    {\phantomsubcaptionlabel{fig-eigenvalues_sketch-e}}
    {\phantomsubcaptionlabel{fig-eigenvalues_sketch-f}}
\end{figure*}

\paragraph{Floquet propagator and dissipative phase transitions}
We consider an open quantum system interacting with a Markovian environment, with time-dependent Hamiltonian and collapse operators.
The dynamics is governed by the Lindblad master equation $\dv{\hat{\rho}}{t} = \mathcal{L} (t) \hat{\rho}$, where
\begin{equation}
    \mathcal{L} (t) \hat{\rho} = -\rmi [\hat{H} (t), \hat{\rho}] + \sum_j \mathcal{D}[\hat{L}_j (t)] \hat{\rho}
\end{equation}
with $\mathcal{D}[\hat{L}_j (t)] \hat{\rho} = \hat{L}_j (t) \hat{\rho} \hat{L}_j^\dagger (t) - \frac{1}{2} \{ \hat{L}_j^\dagger (t) \hat{L}_j (t), \hat{\rho} \}$ the dissipator associated to the collapse operator $\hat{L}_j(t)$.
Throughout this work, we consider a periodic (Floquet) time-dependent superoperator $\mathcal{L} (t)$ (called the Liouvillian), i.e., $\mathcal{L} (t + T) = \mathcal{L} (t)$.

When the Liouvillian is time-independent, its eigenvalues and right eigenvectors are defined by $\mathcal{L}\,\hat{\eta}_j = \lambda_j \hat{\eta}_j$.
The eigenvalues have non-positive real part, $\mathrm{Re}(\lambda_j) \leq 0$, and at least one vanishes, corresponding to the steady state: $\mathcal{L} \hat{\rho}_\mathrm{ss} = 0$, with $\hat{\rho}_\mathrm{ss} \propto \hat{\eta}_0$.
The eigenvalue $\lambda_1$ with the smallest non-zero real part defines the Liouvillian gap, which sets the slowest relaxation rate towards the steady state.
A DPT represents a non-analytic change in the steady state of the system as a function of some control parameter $\zeta$, and it is associated with the closing of the Liouvillian gap, $\Re(\lambda_1) \to 0$, in the thermodynamic limit~\cite{minganti_spectral_2018}.
In this work, we question whether DPTs in time-periodic open quantum systems can be systematically characterized via spectral analysis.

Given the periodicity of the Liouvillian, it is useful to consider the stroboscopic propagator $\mathcal{U}_\mathrm{F} (t^\prime) = \mathcal{U} (t^\prime + T, t^\prime) = \mathcal{T} \exp \qty[ \int_{t^\prime}^{t^\prime + T} \rmd \tau\mathcal{L}(\tau)]$,
a linear, completely positive and trace-preserving map evolving the density matrix by one period $T$.
In some cases, $\mathcal{U}_\mathrm{F} (t^\prime)$ can be associated with an effective time-independent Liouvillian $\mathcal{L}_\mathrm{F}$ such that $\mathcal{U}_\mathrm{F} (t^\prime) = \exp(\mathcal{L}_\mathrm{F} T)$~\cite{minganti_arnoldi-lindblad_2022, campaioli_quantum_2024}.
However, the uniqueness and Markovianity of such a superoperator are not always guaranteed, even for simple periodically driven two-level systems~\cite{schnell_is_2020}.
A rigorous characterization of DPTs in Floquet open quantum systems thus cannot, in general, rely on $\mathcal{L}_\mathrm{F}$, but must instead be grounded in the spectral properties of the propagator $\mathcal{U}_\mathrm{F} (t^\prime)$.
Diagonalizing the $t^\prime$-dependent propagator yields the complex spectrum $\mathcal{U}_\mathrm{F} (t^\prime) \, \hat{\eta}_j^\mathrm{F} (t^\prime) = \varepsilon_j \hat{\eta}_j^\mathrm{F} (t^\prime) = \exp(\lambda_j T) \hat{\eta}_j^\mathrm{F} (t^\prime)$~\cite{Chen2024Periodically}.
Irrespective of the existence of $\mathcal{L}_{\rm F}$, the propagator $\mathcal{U}_\mathrm{F} (t^\prime)$ always admits at least one $t^\prime$-dependent steady state $\hat{\rho}_{\rm ss}^{\rm F} (t^\prime)$ such that $\mathcal{U}_\mathrm{F} (t^\prime) \,\hat{\rho}_{\rm ss}^{\rm F} (t^\prime) = \hat{\rho}_{\rm ss}^{\rm F} (t^\prime)$~\cite{Chen2024Periodically}.
The same spectral information can be accessed through an equivalent, time-independent formulation: in the Sambe space~\cite{Shirley1965, Sambe1973, Bain2001IntroductionFloquetTheory, ChuTelnov2004}, the periodic problem is mapped onto a time-independent Floquet--Liouvillian acting on the system Liouville space enlarged by the $T$-periodic functions, whose eigenvalues are precisely the Floquet exponents $\lambda_j$.
We present this construction and compare the two pictures in the End Matter.

We now define a Floquet DPT between two distinct phases as a non-analytical change of the steady state expectation value of an observable $\hat O$ as an external parameter $\zeta$ approaches some critical value $\zeta_c$.
The DPT is said to be of order $m$ if
\begin{equation}
    \label{eq-dpt}
    \lim_{\zeta\to\zeta_c}\qty|\frac{\partial^m}{\partial\zeta^m}\lim_{N\to\infty} \expval*{\hat{O}}_\mathrm{ss} (\zeta, N)| = +\infty.
\end{equation}
Here $\expval*{\hat{O}}_\mathrm{ss} (\zeta, N) = \frac{1}{T} \int_0^T \rmd t^\prime \operatorname{Tr} [\hat{\rho}_{\rm ss}^{\rm F}(t; \, \zeta,N)\hat{O}]$ is the steady-state expectation value of $\hat{O}$ averaged over one period.
Thus, the non-analytic behavior described by \cref{eq-dpt} corresponds to the closing of the Floquet gap, i.e., the collapse of the leading (non-stationary) eigenvalue onto the unit circle, $|\varepsilon_1| \to 1$, as $N\to\infty$. The position of $\varepsilon_1$ on the unit circle is frame-dependent (see the End Matter).
An illustrative comparison between the gap closure in Liouvillian and Floquet spectra is provided in \cref{fig-eigenvalues_sketch}.

A first-order Floquet DPT occurs if $|\varepsilon_1|=1$ at $\zeta=\zeta_c$ and $\Im(\varepsilon_1)=0$ in its vicinity.
The invariance of $\mathcal{U}_\mathrm{F} (t^\prime)$ under superoperatorial symmetries leads to DPTs with spontaneous symmetry breaking (SSB).
Here, we focus on weak Liouvillian symmetries, associated with a unitary superoperator $\mathcal{S}$ commuting with the dynamical generator, $[\mathcal{S},\,\mathcal{U}_\mathrm{F} (t^\prime)]=0$.
This implies that $\mathcal{U}_\mathrm{F} (t^\prime)$ can be block-diagonalized in symmetry sectors labeled by the eigenvalues $s_n$ of $\mathcal{S}$, that is, $\mathcal{U}_\mathrm{F} (t^\prime) = \bigoplus_{n} \mathcal{U}_\mathrm{F}^{(n)} (t^\prime)$, with $\mathcal{U}_\mathrm{F}^{(n)} (t^\prime) \, \hat{\eta}_{j, n}^{\rm F} (t^\prime) = \varepsilon_{j, n}\,\hat{\eta}_{j, n}^{\rm F} (t^\prime)$ and only the sector with $n=1$ contains the steady state.

A Floquet DPT with SSB occurs when the leading eigenvalue of a symmetry sector reaches the unit circle, $|\varepsilon_{0, n}| \to 1$ (cf. \cref{fig-eigenvalues_sketch-c,fig-eigenvalues_sketch-f}), so that this sector becomes degenerate in modulus with the steady state.
Its position on the unit circle is again frame-dependent (see the End Matter).

In the following, we apply our framework to analyze DPTs in three paradigmatic models of quantum optics: the single- and two-photon driven, dissipative Kerr resonator and the driven quantum Rabi model.
All numerical results have been obtained using the \texttt{QuantumToolbox.jl} package in Julia~\cite{Mercurio2025QuantumToolboxjl}.
The Floquet propagator has been diagonalized using the Arnoldi-Lindblad method~\cite{minganti_arnoldi-lindblad_2022}.

\begin{figure}[t]
    \centering
    \includegraphics{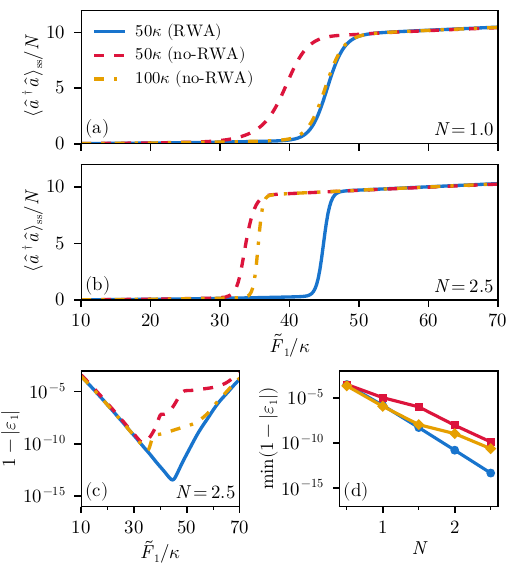}
    \caption{Standard vs Floquet DPT in a single-photon driven, dissipative Kerr oscillator.
    (a,b)~Comparison of the steady-state photon number $\expval*{\hat{a}^\dagger\hat{a}}_\mathrm{ss} / N$ between the RWA approximation and the Floquet picture as a function of $\tilde{F}_1/\kappa$, for $N=1$ and $N=2.5$, respectively.
    (c)~Analysis of the Floquet spectral gap in the vicinity of the DPT as a function of $\tilde F_1 / \kappa$.
    (d)~Scaling of $\min(1-|\varepsilon_1|)$ with the thermodynamic limit of the system. Used parameters: $\omega_0 = 50 \kappa$ (solid blue and dashed red), $\omega_0 = 100 \kappa$ (dash-dotted yellow), $\Delta = -80 \kappa$, $U = 10 \kappa$.}
    \label{fig:1ph-kerr}
    {\phantomsubcaptionlabel{fig:1ph-kerr_photon1}}
    {\phantomsubcaptionlabel{fig:1ph-kerr_photon2.5}}
    {\phantomsubcaptionlabel{fig:1ph-kerr_gap}}
    {\phantomsubcaptionlabel{fig:1ph-kerr_gap_scaling}}
\end{figure}

\paragraph{$n$-photon driven, dissipative Kerr resonators}
Consider a Kerr resonator with an $n$-photon drive described by the time-dependent Hamiltonian
\begin{equation}\label{eq:kerr_full}
    \hat{H} (t) = \omega_0 \hat{a}^\dagger \hat{a} + \frac{U}{2} \hat{a}^{\dagger 2} \hat{a}^2 + 2 F_n \cos (\omega_\mathrm{d} t) \qty( \hat{a}^n + \hat{a}^{\dagger n} ) \, ,
\end{equation}
where $\hat{a}$ ($\hat{a}^\dagger$) is the bosonic annihilation (creation) operator, $\omega_0$ is the resonator frequency, $U$ is the Kerr nonlinearity, $F_n$ is the $n$-photon drive amplitude and $\omega_\mathrm{d}$ is the drive frequency.
The oscillator is subject to both single- and $n$-photon losses, described by the collapse operators $\hat{L}_1 = \sqrt{\kappa}\,\hat{a}$ and $\hat{L}_n = \sqrt{\eta}\,\hat{a}^n$, where $\kappa$ and $\eta$ are the corresponding loss rates.
After applying the RWA on the drive term, the resulting time-independent Hamiltonian reads 
\begin{equation}\label{eq:kerr_rwa}
    \hat{H}_\mathrm{RWA} = \Delta \hat{a}^\dagger \hat{a} + \frac{U}{2} \hat{a}^{\dagger 2} \hat{a}^2 + F_n \qty( \hat{a}^n + \hat{a}^{\dagger n} )\,,
\end{equation}
where $\Delta = \omega_0 - \omega_\mathrm{d}/n$ is the cavity-to-pump detuning.
Notably, both the systems described by \cref{eq:kerr_full,eq:kerr_rwa} and by $\hat{L}_1$ and $\hat{L}_n$, present a weak $\mathbb{Z}_n$ Liouvillian symmetry for $n\ge2$~\cite{minganti_spectral_2018,minganti_dissipative_2023}. 
This implies that the Hamiltonian remains invariant under the transformation $\hat{a} \to \hat{a}\,\rme^{- 2 \rmi \pi / n}$, and the propagator can be block-diagonalized into $n$ symmetry sectors.

The critical properties of the time-independent system for $n=1$ and $n=2$ have been intensively investigated both theoretically and experimentally, from the single resonator model \cite{bartolo_exact_2016, rodriguez_probing_2017, minganti_spectral_2018, fink_signatures_2018, chen_quantum_2023, beaulieu_observation_2025} to chains of nonlinear oscillators \cite{fitzpatrick_observation_2017, biondi_nonequilibrium_2017, vicentini_critical_2018, fedorov_photon_2021, ferrari_chaotic_2025, kruglikov_chaos_2025}.
However, the RWA can give inaccurate predictions in the strong-drive case, with significant deviations from the simple time-independent picture~\cite{Kosata2022Fixing,Seibold2025Floquet}.
We therefore apply our framework to the Floquet DPT of the full time-dependent model in both the single- and two-photon driven cases.

In \cref{fig:1ph-kerr} we study the first-order DPT in the $n=1$ model, approaching the thermodynamic limit by rescaling $U = \tilde{U}/N$ and $F_1 = \tilde{F}_1\sqrt{N}$ for $N\to\infty$~\cite{minganti_spectral_2018}.
\cref{fig:1ph-kerr_photon1,fig:1ph-kerr_photon2.5} show the rescaled steady-state photon number $\expval*{\hat{a}^\dagger\hat{a}}_{\rm ss}/N$ versus the effective drive amplitude $\tilde{F}_1/\kappa$, for two values of $N$: the time-independent RWA model (blue line) and the full time-dependent picture for $\omega_0=50\kappa$ (red-dashed) and $\omega_0=100\kappa$ (yellow-dashdotted).
Both systems exhibit criticality: as $N$ grows, the transition between a vacuum-like and a bright phase sharpens.
However, the time-dependent case shifts the critical point towards smaller $\tilde{F}_1/\kappa$ relative to the RWA prediction.
Interestingly, while $\omega_0=50\kappa$ deviates clearly from the time-independent result already at small $N$, the $\omega_0=100\kappa$ data depart from the RWA only at larger $N$.
This follows because the RWA is valid when $F_1 \ll \omega_0,\,\omega_\mathrm{d}$: as the thermodynamic limit is approached, the scaling of $F_1$ with $N$ eventually breaks down the approximation, even for large but finite $\omega_0/\kappa$.

\begin{figure}[t]
    \centering
    \includegraphics{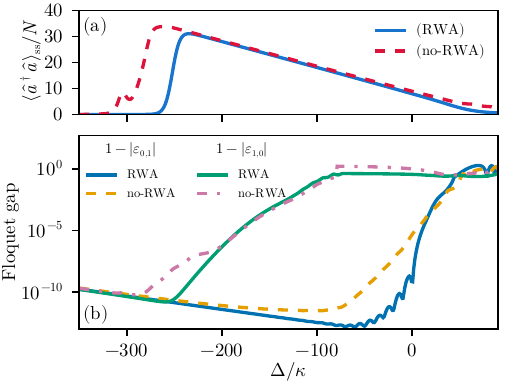}
    \caption{Standard vs Floquet DPT in a two-photon driven, dissipative Kerr oscillator.
    (a)~Comparison between the steady-state photon number $\expval*{\hat a^\dagger \hat a}_{\rm ss}/N$ as a function of $\Delta/\kappa$ in the RWA approximation and within the time-dependent picture.
    (b)~Analysis of the Floquet spectral gap in both symmetry sectors as a function of $\Delta/\kappa$. Used parameters: $\omega_0 = 100 \kappa$, $U = 10 \kappa$, $\eta = \kappa / 2$, $F_2 = 40 \kappa$, and $N = 1$.}
    \label{fig:2ph-kerr}
    {\phantomsubcaptionlabel{fig:2ph-kerr_photon}}
    {\phantomsubcaptionlabel{fig:2ph-kerr_gap}}
\end{figure}

The time dependence not only shifts the critical point but also modifies the Floquet gap, which we quantify through the distance of the leading non-stationary eigenvalue from the unit circle, $1-|\varepsilon_1|$.
\cref{fig:1ph-kerr_gap} shows that the RWA underestimates the spectral gap by several orders of magnitude, so that the critical slowing down~\cite{vicentini_critical_2018,beaulieu_observation_2025} due to dissipative criticality is much less pronounced once the counter-rotating terms are included.
Importantly, this difference grows as the thermodynamic limit is approached, regardless of $\omega_0/\kappa$ [\cref{fig:1ph-kerr_gap_scaling}].
Collectively, these results show that the intrinsic time dependence changes the scaling towards the thermodynamic limit, and that the RWA fails to capture the critical slowing down in the critical regime.

In \cref{fig:2ph-kerr} we analyze the second-order DPT of the $n=2$ model, where the thermodynamic limit follows from rescaling $U = \tilde{U}/N$ as $N\to\infty$~\cite{minganti_spectral_2018}. \cref{fig:2ph-kerr_photon} shows the steady-state photon number $\expval*{\hat{a}^\dagger\hat{a}}_{\rm ss}/N$ versus $\Delta/\kappa$.
As in the single-photon case, the time-dependent model shifts the critical point and increases the steady-state population in the bright phase compared to the RWA.
The properties of the Floquet gap are presented in \cref{fig:2ph-kerr_gap}, where we observe its closing, $|\varepsilon_1| \to 1$, as the leading eigenvalues of the two $\mathbb{Z}_2$ symmetry sectors reach the unit circle. In the lab frame, the odd-sector eigenvalue approaches $\varepsilon_{0,1} \to -1$: the $\mathbb{Z}_2$-odd order parameter $\expval*{\hat{a}}$ then responds at the subharmonic $\omega_\mathrm{d}/2$, a period-doubling of the symmetry-broken states that links this transition to dissipative discrete time crystals~\cite{else_floquet_2016, gong_discrete_2018, lazarides_time_2020,RieraCampeny2020timecrystallinity}.
Again, the RWA underestimates the Liouvillian gap by several orders of magnitude, and thus overestimates the timescale of the critical slowing down near the transition.

\paragraph{Quantum Rabi model}
The quantum Rabi model (QRM) is the paradigmatic example of light-matter interaction, describing the coupling between a two-level system (TLS) and a single bosonic mode~\cite{Rabi1936OnTheProcess,Braak2011Integrability}. The Hamiltonian is given by
\begin{equation}
    \label{eq:QRM-hamiltonian}
    \hat{H}_\mathrm{QRM} = \omega_{\rm c} \hat{a}^\dagger \hat{a} + \frac{\omega_\mathrm{q}}{2} \hat{\sigma}_z + g (\hat{a} + \hat{a}^\dagger) \hat{\sigma}_x \, ,
\end{equation}
where $\hat{a}$ ($\hat{a}^\dagger$) is the bosonic annihilation (creation) operator, $\hat{\sigma}_z$ and $\hat{\sigma}_x$ are the Pauli operators for the TLS, $\omega_{\rm c}$ is the frequency of the bosonic mode, $\omega_\mathrm{q}$ is the transition frequency of the TLS and $g$ is the coupling strength. The Hamiltonian in \cref{eq:QRM-hamiltonian} describes a wide range of systems in cavity and circuit quantum electrodynamics~\cite{Niemczyk2010Circuit,FornDiaz2017Ultrastrong,Yoshihara2017Superconducting,Yoshihara2018Inversion,FriskKockum2019Ultrastrong,FornDiaz2019Ultrastrong,DeBernardis2024Tutorial, lamberto_renormalization_2025}, with a rich phenomenology including deterministic nonlinear optics effects~\cite{Ma2015ThreePhoton,Garziano2016OnePhoton,Kockum2017Deterministic,Macr2022,Carlo2022s, kockum_frequency_2017, wang_strong_2025}, gauge invariance and photodetection issues~\cite{DeBernardis2018Breakdown, stokes_gauge_2019, DiStefano2019Resolution,Savasta2021Gauge, di_stefano_photodetection_2018, le_boite_theoretical_2020, mercurio_regimes_2022}, as well as quantum phase transitions~\cite{Ashhab2013Superradiance,Hwang2015Quantum,Hwang2018Dissipative,Lyu2024Multicritical,Xiao2026Metastability}.

For $g \ll \omega_{\rm c},\,\omega_\mathrm{q}$, the RWA neglects the counter-rotating terms $\hat{\sigma}_+ \hat{a}^\dagger$ and $\hat{\sigma}_- \hat{a}$, leading to the Jaynes-Cummings model (JCM)~\cite{Jaynes1963Comparison} $\hat{H}_\mathrm{JCM} = \omega_{\rm c} \hat{a}^\dagger \hat{a} + \frac{\omega_\mathrm{q}}{2} \hat{\sigma}_z + g (\hat{\sigma}_+ \hat{a} + \hat{\sigma}_- \hat{a}^\dagger)$, which is exactly solvable and widely used to describe light-matter interaction in the weak and strong coupling regimes~\cite{Wineland2013Nobel,Haroche2013Nobel}. The JCM is known to exhibit a DPT, related to the breakdown of the photon blockade~\cite{Carmichael2015Breakdown,fink_observation_2017, GutierrezJauregui2018Dissipative,Vukics2019Finitesize}.
However, the RWA breaks down in the ultrastrong coupling (USC) regime, where $g$ becomes comparable to $\omega_{\rm c}$ and $\omega_\mathrm{q}$, and the full QRM must be used.
When $g$ exceeds $\omega_{\rm c}$ and $\omega_{\rm q}$, the QRM enters the deep strong coupling (DSC) regime, where the system exhibits light-matter decoupling~\cite{DeLiberato2014LightMatter, garcia-ripoll_light-matter_2015,Mueller2020Deep, Ashida2021Cavity, mercurio_regimes_2022}: the effective mass of the dressed TLS diverges, suppressing the light-matter interaction and collapsing the energy spectrum into degenerate pairs.
Two questions thus arise: how the USC regime modifies the DPT, and whether the transition persists in the DSC regime.

\begin{figure}[b]
    \centering
    \includegraphics{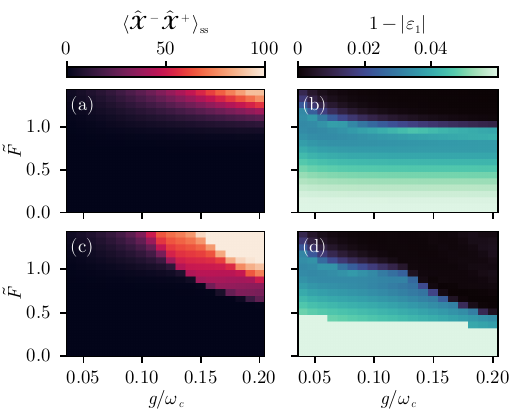}
    \caption{Driven-dissipative JCM (a,b) and QRM (c,d) from weak to ultrastrong coupling. Panels (a,c) show the steady-state output field field versus coupling $g$ and rescaled drive $\tilde F$, while panels (b,d) report the corresponding Floquet gap $1-|\varepsilon_1|$. Used parameters are $\omega_\mathrm{c} = \omega_{\rm q} = \omega_{\rm d} = 50 \kappa$.}
    \label{fig:jc_qrm_weak_to_usc}
    {\phantomsubcaptionlabel{fig:jc_qrm_weak_to_usc--JC_population}}
    {\phantomsubcaptionlabel{fig:jc_qrm_weak_to_usc--JC_gap}}
    {\phantomsubcaptionlabel{fig:jc_qrm_weak_to_usc--QRM_population}}
    {\phantomsubcaptionlabel{fig:jc_qrm_weak_to_usc--QRM_gap}}
\end{figure}

First, we consider the driven-dissipative JCM, where the total Hamiltonian in the drive frame and within the RWA is given by $\hat{H}_\mathrm{JCM}^\mathrm{tot} = \hat{H}_\mathrm{JCM} - \omega_\mathrm{d} \hat{a}^\dagger \hat{a} - \omega_\mathrm{d} \hat{\sigma}_z / 2 + F (\hat{a} + \hat{a}^\dagger)$.  
The system is subject to single-photon losses with collapse operator $\hat{L} = \sqrt{\kappa}\,\hat{a}$, where $\kappa$ is the loss rate. The driven-dissipative JCM then exhibits a second-order DPT at a critical drive amplitude $F_c = g / 2$~\cite{Carmichael2015Breakdown}, as shown in \cref{fig:jc_qrm_weak_to_usc--JC_population,fig:jc_qrm_weak_to_usc--JC_gap}, where the steady-state photon number and Floquet gap are plotted versus the coupling strength $g$ and the rescaled drive amplitude $\tilde{F} = 2 F / g$.

Next, we study the driven-dissipative QRM, with Hamiltonian
\begin{equation}\label{eq:driven_qrm}
    \hat{H}_\mathrm{QRM}^\mathrm{tot} (t) = \hat{H}_\mathrm{QRM} + F (\hat{X}^+ e^{\rmi \omega_\mathrm{d} t} + \hat{X}^- e^{-\rmi \omega_\mathrm{d} t})\, ,
\end{equation}
where $\hat{X}^+ = \sum_{j, k>j} \mel{j}{\rmi (\hat{a} - \hat{a}^\dagger)}{k} \dyad{j}{k}$ and $\hat{X}^- = (\hat{X}^+)^\dagger$ are the positive and negative frequency components of the quadrature operator, respectively, and $\ket{j}$ are the eigenstates of $\hat{H}_\mathrm{QRM}$~\cite{Ridolfo2012Photon,LeBoite2016Fate,le_boite_theoretical_2020}.
Note that in \cref{eq:driven_qrm} we are deliberately neglecting the counter-rotating terms in the drive in order to isolate the effect of the counter-rotating terms in the light-matter interaction and compare it to the driven JCM.

\begin{figure}[t]
    \centering
    \includegraphics{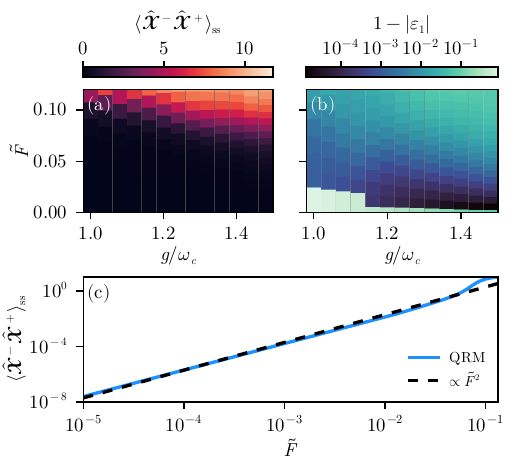}
    \caption{Driven-dissipative QRM in the deep-strong-coupling regime. (a) Steady-state output field versus $g$ and $\tilde F$. (b) Floquet gap $1-|\varepsilon_1|$, showing suppression of criticality as $g$ increases. (c) Output field scaling at fixed coupling, consistent with a quadratic dependence on $\tilde F$ as expected for a driven-dissipative harmonic oscillator. Parameters as in \cref{fig:jc_qrm_weak_to_usc}.}
    \label{fig:qrm_dsc}
    {\phantomsubcaptionlabel{fig:qrm_dsc--population}}
    {\phantomsubcaptionlabel{fig:qrm_dsc--gap}}
    {\phantomsubcaptionlabel{fig:qrm_dsc--population_trend}}
\end{figure}

In the USC and DSC regimes, the open-system evolution can no longer be described by a standard Lindblad master equation, but requires a more general model accounting for the strong hybridization between light and matter degrees of freedom~\cite{Beaudoin2011Dissipation,Settineri2018Dissipation, mercurio_regimes_2022, Mercurio2023PureDephasing}.
The resulting master equation is governed by a time-periodic Liouvillian $\mathcal{L}_\mathrm{GME} (t)$, detailed in the End Matter.
In \cref{fig:jc_qrm_weak_to_usc--QRM_population,fig:jc_qrm_weak_to_usc--QRM_gap} we plot the steady-state output field $\expval*{\hat{\mathcal{X}}^- \hat{\mathcal{X}}^+}_\mathrm{ss}$ with $\hat{\mathcal{X}}^+ = \frac{1}{\omega_{\rm c}} \frac{\rmd}{\rmd t} \hat{X}^+ = - i \sum_{j, k>j} \frac{\omega_k - \omega_j}{\omega_{\rm c}} \mel{j}{\rmi (\hat{a} - \hat{a}^\dagger)}{k} \dyad{j}{k}$~\cite{Ridolfo2012Photon,LeBoite2016Fate,XRWAOutputFieldNote} and the Floquet gap $1-|\varepsilon_1|$ as a function of $g$ and $\tilde{F}$ for the driven-dissipative QRM. 
The phase transition shifts towards smaller $\tilde{F}$ as we enter the USC regime, a behavior entirely missed by the JCM.

Finally, \cref{fig:qrm_dsc--population,fig:qrm_dsc--gap} show the steady-state output field and the Floquet gap in the DSC regime.
The critical point converges to zero as $g$ increases, and the DPT disappears.
This is a consequence of light-matter decoupling, which prevents the system from undergoing the transition.
This finding demonstrates that counter-rotating terms in the light-matter interaction have a remarkable impact on the critical properties, inducing both quantitative and qualitative differences with respect to the simpler RWA model.
In more detail, \cref{fig:qrm_dsc--population_trend} plots the steady-state output field versus $\tilde{F}$ for $g = 1.5 \,\omega_{\rm c}$, revealing a $\tilde{F}^2$ dependence, as for a single uncoupled driven-dissipative harmonic oscillator.

\paragraph{Discussion}
We introduced a general spectral framework to characterize DPTs in periodically driven open quantum systems, based on the Floquet propagator spectrum. 
This analysis shows that counter-rotating terms can modify or even suppress critical behavior. 
Quantitatively, they can induce substantial changes in the propagator gap, with direct consequences for relaxation time scales.
Qualitatively, they can eliminate the transition altogether, as demonstrated for the driven QRM in the DSC limit.
These results indicate that Floquet spectral methods are not only formally required, but also relevant in realistic regimes where RWA break down near criticality. More broadly, our approach opens the way to investigate dissipative critical phenomena in systems where time dependence cannot be neglected, such as parametrically driven cavity- and circuit-QED settings.

\begin{acknowledgments}
\paragraph{Acknowledgements}
We acknowledge enlightening discussions with Daniele Lamberto.
A.M., F.F., L.F. and V.S. acknowledge support from the Swiss National Science Foundation through Projects No. 200020\_215172, 200021-227992, and 20QU-1\_215928, and as a part of NCCR SPIN (grant number 225153).
A.M. acknowledges the Swiss Quantum Initiative (SQI) of the Swiss Academy of Sciences (SCNAT) through the 2024 Quantum Voucher Model for the Grant/Ruling 24\_1084. V.M. acknowledge PNRR MUR project, National Quantum Science and Technology Institute (NQSTI) Grant No. PE0000023.
\end{acknowledgments}

\bibliography{biblio}

\vspace{2cm}

\onecolumngrid
\begin{center}
  \textbf{\large End Matter}\\[1em]
\end{center}
\vspace{1em}

\twocolumngrid

\paragraph{Comparison with Sambe space approach}
The Floquet propagator spectrum employed in the main text admits an equivalent, fully time-independent formulation in the so-called Sambe (or Floquet--Liouville) space~\cite{Shirley1965, Sambe1973, Bain2001IntroductionFloquetTheory, ChuTelnov2004}.
By Floquet's theorem, the solutions of the periodic Lindblad equation can be written as $\hat{\rho}_j (t) = \rme^{\lambda_j t} \hat{\eta}_j (t)$, with a periodic $\hat{\eta}_j (t + T) = \hat{\eta}_j (t)$.
Inserting this ansatz into the master equation turns the time evolution into the time-independent eigenvalue problem
\begin{equation}
    \label{eq:sambe-eigenproblem}
    \qty[\mathcal{L} (t) - \partial_t] \hat{\eta}_j (t) = \lambda_j \, \hat{\eta}_j (t) \, ,
\end{equation}
defined on the extended space $\mathcal{F} = \mathcal{B} (\mathcal{H}) \otimes L^2_T$, where $\mathcal{B} (\mathcal{H})$ is the Liouville space of operators acting on the system Hilbert space $\mathcal{H}$, of dimension $d^2$ with $d = \dim \mathcal{H}$, and $L^2_T$ is the Hilbert space of square-integrable $T$-periodic functions spanned by $\{ \rme^{\rmi n \omega t} \}_{n \in \mathbb{Z}}$, with $\omega = 2 \pi / T$ and inner product $\langle\!\langle f | g \rangle\!\rangle = \frac{1}{T} \int_0^T f^* (t) g (t) \, \rmd t$.

Expanding the Liouvillian and the eigenoperators in Fourier series, $\mathcal{L} (t) = \sum_k \mathcal{L}^{(k)} \rme^{\rmi k \omega t}$ and $\hat{\eta}_j (t) = \sum_n \hat{\eta}_j^{(n)} \rme^{\rmi n \omega t}$, the Sambe-space Liouvillian $\mathcal{L}_\mathrm{S} \equiv \mathcal{L} (t) - \partial_t$ takes the block form
\begin{equation}
    \label{eq:sambe-matrix}
    \begin{split}
            [\mathcal{L}_\mathrm{S}]_{mn} = & \, \mathcal{L}^{(m - n)} - \rmi m \omega \, \delta_{mn} \, , \\ 
            \mathcal{L}^{(k)} = & \, \frac{1}{T} \int_0^T \mathcal{L} (t) \, \rme^{-\rmi k \omega t} \, \rmd t \, .
    \end{split}
\end{equation}
For a monochromatic drive, only $\mathcal{L}^{(0)}$ and $\mathcal{L}^{(\pm 1)}$ are nonzero, and $\mathcal{L}_\mathrm{S}$ is block-tridiagonal.
For numerical diagonalization, the formally infinite Fourier ladder must be truncated to a finite cutoff $|n| \le k_\mathrm{max}$, so that $\mathcal{L}_\mathrm{S}$ becomes a matrix of dimension $d^2 (2 k_\mathrm{max} + 1)$; the exact result is recovered in the limit $k_\mathrm{max} \to \infty$.

The Sambe and propagator pictures are exactly equivalent, with eigenvalues related by
\begin{equation}
    \label{eq:sambe-propagator}
    \varepsilon_j = \rme^{\lambda_j T} \, .
\end{equation}
The Sambe spectrum is redundant: if $\hat{\eta}_j (t)$ is an eigenoperator with eigenvalue $\lambda_j$, then $\hat{\eta}_j (t) \, \rme^{-\rmi n \omega t}$ is also one, with eigenvalue $\lambda_j + \rmi n \omega$.
All such replicas map onto the same $\varepsilon_j$ because $\rme^{\rmi n \omega T} = 1$, so that folding this Floquet--Brillouin-zone redundancy and exponentiating recovers the unit-disk spectrum of $\mathcal{U}_\mathrm{F} (t^\prime)$.
Accordingly, a Floquet DPT corresponds to the closing of the Floquet--Liouvillian gap, $\Re (\lambda_1) \to 0$, in direct analogy with the time-independent criterion~\cite{minganti_spectral_2018}.
We emphasize that $\mathcal{L}_\mathrm{S}$ should not be confused with the effective generator $\mathcal{L}_\mathrm{F} = \frac{1}{T} \log \mathcal{U}_\mathrm{F}$ mentioned in the main text: the former is always well defined, the branch ambiguity of the logarithm corresponding here to the choice of Brillouin-zone representative.

\begin{figure}[t]
    \centering
    \includegraphics{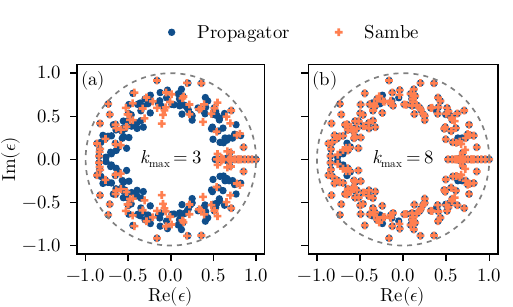}
    \caption{Convergence of the Sambe-space spectrum to the Floquet propagator spectrum for the single-photon driven, dissipative Kerr resonator. Propagator eigenvalues $\varepsilon_j$ (blue dots) and exponentiated Sambe eigenvalues $\rme^{\lambda_j T}$ (orange crosses). (a)~For a small Fourier cutoff $k_\mathrm{max} = 3$, the truncated Sambe spectrum still deviates from the propagator one. (b)~For $k_\mathrm{max} = 8$, the two spectra coincide, confirming the equivalence of the two approaches. Parameters as in \cref{fig:1ph-kerr} with $\omega_0 = 50\kappa$, $N = 1.9$ and $\tilde{F} = 5\kappa$.}
    \label{fig:sambe_comparison}
    {\phantomsubcaptionlabel{fig:sambe_comparison-a}}
    {\phantomsubcaptionlabel{fig:sambe_comparison-b}}
\end{figure}

Both our formulation in terms of the Floquet propagator, and the one in terms of $\mathcal L_\mathrm{S}$ are thus rigorous and equivalent.
We adopt the propagator throughout because it is finite-dimensional, acting on $\mathcal{B} (\mathcal{H})$ rather than on the enlarged space $\mathcal{F}$, and free of any frequency truncation. It is therefore numerically robust in the strongly driven regimes, with large amplitudes and non-negligible counter-rotating terms, where the Sambe expansion requires many harmonics to converge.
\cref{fig:sambe_comparison} confirms this equivalence for the single-photon Kerr resonator: as the Fourier cutoff $k_\mathrm{max}$ is increased, the exponentiated Sambe eigenvalues $\rme^{\lambda_j T}$ converge to the propagator eigenvalues $\varepsilon_j$, still deviating at $k_\mathrm{max} = 3$ (\cref{fig:sambe_comparison-a}) but becoming almost indistinguishable at $k_\mathrm{max} = 8$ (\cref{fig:sambe_comparison-b}).

\paragraph{Frame dependence of the Floquet spectrum}
The eigenvalues $\varepsilon_j$ of the one-period propagator $\mathcal{U}_\mathrm{F}(t^\prime)$ are invariant under any change of frame $\hat{\rho}\to\hat{V}(t)\hat{\rho}\hat{V}^\dagger(t)$ that is itself $T$-periodic, $\hat{V}(t+T)=\hat{V}(t)$. A non-$T$-periodic frame with $\hat{V}(t+T)=\hat{W}\hat{V}(t)$ instead yields a genuinely different one-period map. When $\hat{W}$ commutes with $\mathcal{U}_\mathrm{F}(t^\prime)$, propagator and twist share their eigenmodes, and each eigenvalue is multiplied by the corresponding unimodular eigenvalue of $\hat{W}$. In the case of the $\mathbb{Z}_n$ symmetry, the moduli $|\varepsilon_j|$ are preserved and only the phases $\arg(\varepsilon_j)$ are shifted.

The two-photon Kerr resonator illustrates this concretely. The frame that renders the parametric drive static rotates at $\omega_\mathrm{d}/2$, $\hat{V}(t)=\exp[\rmi(\omega_\mathrm{d}/2)\,t\,\hat{a}^\dagger\hat{a}]$, and obeys $\hat{V}(t+T)=\hat{V}(t)\,\rme^{\rmi\pi\hat{a}^\dagger\hat{a}}=\hat{V}(t)\,\mathcal{S}$: it is twisted by the $\mathbb{Z}_2$ parity $\mathcal{S}$ each period (and is $2T$-periodic). It therefore maps the odd-sector eigenvalue from $\varepsilon_{0,1}\to-1$ in the lab frame onto $\varepsilon_{0,1}\to+1$ in the drive frame. By contrast, for the single-photon drive ($n=1$) the analogous frame rotates at $\omega_\mathrm{d}$ and is genuinely $T$-periodic, so no phase shift occurs and $\varepsilon_1\to+1$ in every frame.

\paragraph{Time-resolved simulations}
The dramatic difference in the Floquet gap between the RWA and the full time-dependent model reported in the main text (cf. \cref{fig:1ph-kerr_gap,fig:1ph-kerr_gap_scaling}) has a direct, observable consequence on the relaxation dynamics.
Since the gap sets the slowest rate at which the system approaches its steady state, a larger gap translates into a faster relaxation and, conversely, a smaller gap into a more pronounced critical slowing down.
To make this explicit, we compare the time-resolved evolution of the mean photon number $\expval*{\hat{a}^\dagger\hat{a}} (t)$ for the single-photon driven Kerr resonator, evolving the system from the vacuum under both the RWA Hamiltonian [\cref{eq:kerr_rwa}] and the full time-dependent one [\cref{eq:kerr_full}].

\begin{figure}[b]
    \centering
    \includegraphics{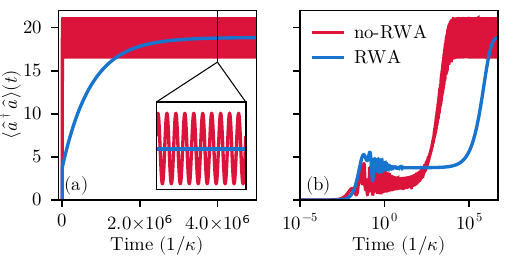}
    \caption{Time-resolved relaxation towards the steady state of the single-photon driven, dissipative Kerr resonator, comparing the RWA model [\cref{eq:kerr_rwa}, blue] with the full time-dependent model [\cref{eq:kerr_full}, red].
    The mean photon number $\expval*{\hat{a}^\dagger\hat{a}} (t)$ is evolved from the vacuum at a fixed drive amplitude in the bright phase, close to the critical point.
    (a) Linear time scale: the RWA relaxes slowly and monotonically, reaching its steady state only after $\sim 10^{6}\,\kappa^{-1}$, whereas the full model settles orders of magnitude earlier.
    (b) The same data on a logarithmic time axis, making the separation of relaxation time scales explicit.
    Parameters as in \cref{fig:1ph-kerr} with $\omega_0 = 50\kappa$, $N = 1.9$ and $\tilde{F} = 55\kappa$.}
    \label{fig:time_resolved_1ph-kerr}
\end{figure}

The results are shown in \cref{fig:time_resolved_1ph-kerr}.
The RWA dynamics (blue) relaxes slowly and monotonically, reaching its steady-state value only after about $10^6\,\kappa^{-1}$.
By contrast, the full model (red) settles almost immediately, on a time scale orders of magnitude shorter, in agreement with its much larger Floquet gap.
Once relaxed, the full model does not approach a static value but a time-periodic steady state: the red curve oscillates rapidly within a narrow band at the drive frequency (micromotion), as highlighted in the inset, whereas the RWA steady state is time-independent.
This confirms that, in the vicinity of the transition, the RWA largely overestimates the critical slowing down, while the counter-rotating terms substantially accelerate the approach to the steady state, in full agreement with the spectral-gap analysis of the main text.

\paragraph{Generalized master equation}
In order to describe the driven-dissipative QRM at arbitrary coupling strengths, we need to go beyond the standard Lindblad master equation and consider a more general model that accounts for the strong hybridization between light and matter degrees of freedom~\cite{Beaudoin2011Dissipation,Settineri2018Dissipation, mercurio_regimes_2022, Mercurio2023PureDephasing}. Although a dressed master equation in the dressed basis can be derived~\cite{Beaudoin2011Dissipation}, it fails in the presence of degeneracies or harmonic spectra, as in the weak or DSC regimes~\cite{mercurio_regimes_2022}. To this end, we use a generalized master equation (GME) approach. The density matrix evolves according to a time-periodic Liouvillian $\mathcal{L}_\mathrm{GME} (t)$, which at zero temperature can be expressed as
\begin{equation}
    \label{eq:GME}
    \mathcal{L}_\mathrm{GME} (t) = \mathcal{L}_0 + \mathcal{L}_{+1} {\exp} (\rmi \omega_\mathrm{d} t) + \mathcal{L}_{-1} {\exp} (-\rmi \omega_\mathrm{d} t) \, ,
\end{equation}
with $\mathcal{L}_{\pm 1} \hat\rho = -\rmi \comm{\hat{X}^{\pm}}{\hat{\rho}}$ being the positive and negative frequency components of the drive term, and $\mathcal{L}_0$ given by
\begin{equation}
    \label{eq:GME-L0}
    \begin{split}
        \mathcal{L}_0 \hat{\rho} =& \, -\rmi \comm{\hat{H}_\mathrm{QRM}}{\hat{\rho}} + \frac{1}{2} 
        \sum_{\substack{j, k>j \\ l, m>l}} \tilde{\kappa} (\omega_{ml}) \times \\
        & \left[ \hat{X}_{lm} \hat{\rho} \hat{X}_{jk}^{\dagger} - \hat{X}_{jk}^{\dagger} \hat{X}_{lm} \hat{\rho} + \hat{X}_{jk} \hat{\rho} \hat{X}_{lm}^{\dagger} - \hat{\rho} \hat{X}_{lm}^{\dagger} \hat{X}_{jk} \right] \, .
    \end{split}
\end{equation}
Here $\omega_{ml} = \omega_m - \omega_l$ is the transition frequency between the eigenstates $\ket{m}$ and $\ket{l}$ of $\hat{H}_\mathrm{QRM}$, $\hat{X}_{jk} = \mel{j}{\rmi (\hat{a} - \hat{a}^\dagger)}{k} \dyad{j}{k}$, and $\tilde{\kappa} (\omega) = \kappa \, \omega / \omega_{\rm c}$ is the spectral density of an Ohmic bath. In the weak coupling regime, the GME reduces to the standard Lindblad master equation with a single collapse operator $\hat{L} = \sqrt{\kappa}\,\hat{a}$.

\end{document}